\documentclass[aps,12pt,superscriptaddress,amsfonts,amssymb,amsmath]{revtex4}
\pdfoutput=1

\usepackage{graphicx}
\usepackage{epsfig}
\usepackage{epstopdf}
\usepackage{natbib}
\usepackage{xcolor}

\usepackage{verbatim}
\usepackage{empheq}
\usepackage{dsfont}
\usepackage{caption}
\usepackage{subcaption}
\usepackage{tabulary}
\usepackage{nccmath}
\usepackage{flafter}

\usepackage{array}

\newcolumntype{F}[1]{%
    >{\raggedright\arraybackslash\hspace{0pt}}p{#1}}%
\newcolumntype{T}[1]{%
    >{\centering\arraybackslash\hspace{0pt}}p{#1}}%

\begin{document}

\title{The Bass diffusion model: agent-based implementation on arbitrary networks}

\author{L.\ Di Lucchio \footnote{Email address: Laura.DiLucchio@unibz.it}}
\affiliation{Free University of Bozen-Bolzano \\ Faculty of Engineering \\ I-39100 Bolzano, Italy}

\author{G.\ Modanese \footnote{Email address: Giovanni.Modanese@unibz.it}}
\affiliation{Free University of Bozen-Bolzano \\ Faculty of Engineering \\ I-39100 Bolzano, Italy}
\date{\today}

\linespread{0.9}

\begin{abstract}
We show how the combined use of the free software packages \texttt{networkX} and \texttt{NetLogo} allows to implement quickly and with large flexibility agent-based network simulations of the classical Bass diffusion model and of its extensions and modifications. In addition to the standard internal graph implementations available in \texttt{NetLogo} (random, Barabasi-Albert-1 and small world), one can thus employ more complex Barabasi-Albert and small-world networks, plus scale-free networks with arbitrary power-law exponent $\gamma$ built in \texttt{networkX} through a configuration model algorithm. It is also possible to induce degree correlations in the networks in a controlled way via Newman rewiring and to simulate dynamics on arbitrary signed networks (networks where link can have positive or negative weights, with corresponding effects on diffusion). Some new results obtained in the agent-based simulations (and differing from those in mean-field approximation) are the following. The introduction of assortative correlations in scale-free networks has the effect of delaying the adoption peak in the Bass model, compared to the uncorrelated case. The peak time $t_{max}$ depends strongly also on the maximum degree effectively present in the network. For diffusion models with threshold on signed network, if negative influences have a weight equal to or greater than positive influences, then a high level of clustering tends to cause adoption blockades. In connection to this, by analysing statistical ensembles of assortative scale-free networks generated via Newman rewiring one observes a remarkable strong correlation between the function of the average degree of first neighbors $\bar{k}_{nn}(k)$ and the average clustering coefficient depending on the degree $\bar{C}(k)$.

\end{abstract}

\maketitle

\section{Introduction}

In the quantitative analysis of complex systems, in particular for applications to social science and economics, agent-based simulations represent a valid alternative and complement to modelling techniques based on ordinary or stochastic differential equations. Agent-based simulations benefit from the steady increase in available computational power. They allow to represent the behavior of single individuals and to discover the dynamical emergence of collective behavior and interactions at an aggregate level.

We are mainly interested in this work into agent-based diffusion models of innovation and technological change, a phenomenon of great economic and social relevance, systematically studied since the 1960s \cite{rogers2010diffusion} and mathematically formalized through the Bass equation and its developments (see e.g. \cite{bass1969new,norton1987diffusion,jiang2012generalized,guidolin2023innovation}). The review by Kiesling et al.\ \cite{kiesling2012agent} summarizes briefly the classical theory of Bass and describes all the main agent-based models of innovation diffusion, incorporating contents of previous reviews \cite{garcia2005uses,dawid2006agent,peres2010innovation}. Kiesling et al.\ emphasize the importance of agent-based models for overcoming issues and limitations of the analytical models through an individual-based modeling approach.

More specifically, we will consider the implementation of diffusion models on complex networks. This is an essential aspect for social science models, independently from their detailed formulation. In our previous work we investigated the Bass equation on many kinds of scale-free networks in Heterogeneous Mean Field approximation \cite{bertotti2016innovation,bertotti2016bass,bertotti2019bass}. Recently we developed some numerical techniques, based on Newman rewiring, for the explicit construction of scale-free networks with assortative degree correlations (a typical feature of social networks). On these networks it is possible to solve the Bass equations for each node. The codes were first written in \texttt{C++} and \texttt{Mathematica}, then in \texttt{Python}, taking full advantage of object-based programming with the powerful package \texttt{networkX} \cite{di2023generation,hagberg2008exploring,platt2019network,son2023revisiting}.

\begin{figure}[ht] 
\centering \includegraphics[width=0.6\columnwidth]{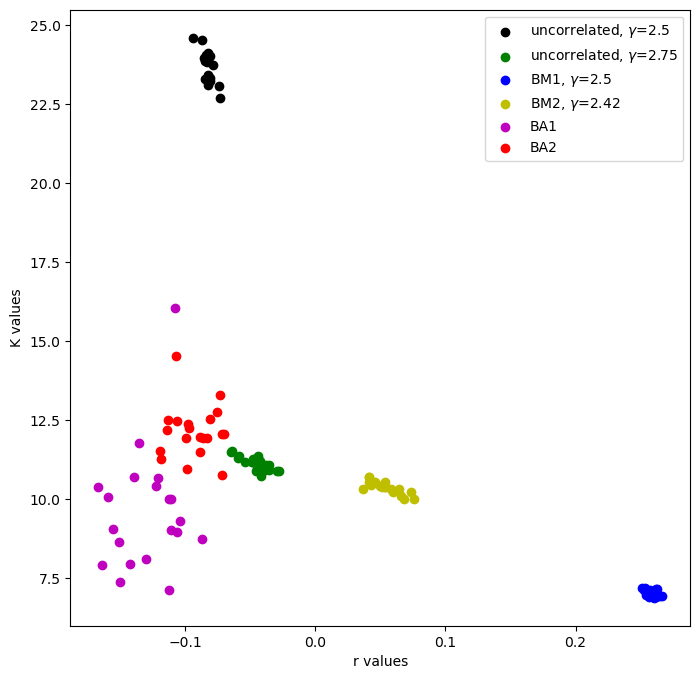}
\caption{
Location in the $r$-$K$ plane of some small statistical ensembles of scale-free networks, each one with 1000 nodes. The Barabasi-Albert-1 and Barabasi-Albert-2 networks are generated with a standard preferential attachment process, in which new nodes are attached respectively to 1 and 2 existing nodes. Their maximum degree is variable, which accounts for the larger dispersion. In the other ''clouds'' the degree distribution is fixed through a configuration model algorithm; the different realizations have been obtained by Newman rewiring with assortative target correlations for BM1 and BM2 \cite{di2023generation}, and with uncorrelated target correlations for the remaining ones. The plot coordinates $r$ and $K$ represent respectively the Newman assortativity coefficient and the network average of the nearest-neighbor degrees of the nodes (see exact definition in the text, and also see the general relation between $\Delta K$ and $\Delta r$, according to which in a degree-conserving rewiring $\Delta K$ and $\Delta r$ always have opposite signs).
}
\label{plot-r-k}
\end{figure}

\begin{figure}[ht] 
\centering \includegraphics[width=0.6\columnwidth]{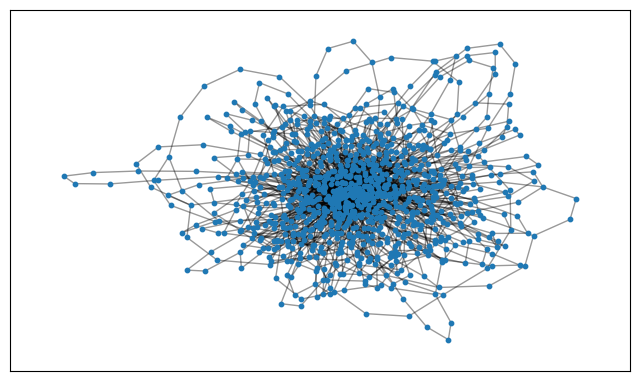}
\caption{
An example of an assortative network obtained via Newman rewiring from a scale-free configuration model network with 1000 nodes and critical exponent $\gamma=2.5$. Typical values of $r$ and $K$ for networks of this kind are shown in Fig.\ \ref{plot-r-k}. The minimum degree of the nodes is 2 and their average degree about 4. Note the strongly connected central region, where the hubs link preferentially to other hubs (compare Fig.\ \ref{knn-C-ASS}). The long chains in the periphery are due to the probabilistic enforcement in the rewiring of a large value of the correlation $P(2|2)$.
}
\label{assort-net}
\end{figure}

Some typical features of the networks generated by Newman rewiring are represented in Figs.\ \ref{plot-r-k}, \ref{assort-net}. Fig.\ \ref{plot-r-k} shows an $r$-$K$ plot of some small network ensembles generated with different target correlations and, for comparison, two ''clouds'' of Barabasi-Albert networks built with a standard stochastic preferential attachment process. The quantities $r$ and $K$ represent respectively the Newman assortativity coefficient and the network average of the average nearest neighbor degree of each node. 
In terms of the degree distribution $P(k)$ and of the average nearest neighbor degree function $\bar{k}_{nn}(k)$, one has $K=\sum_k P(k)\bar{k}_{nn}(k)$.
The use of $r$-$K$ plots is quite natural for assortative and disassortative rewiring. When the rewiring preserves the degree distribution, the variations $\Delta r$ and $\Delta K$ in a rewiring step are univocally related by
\begin{equation}
    \Delta K = \frac{L\sigma_q^2}{2N} \frac{(AD+BC)}{ABCD} \Delta r
\end{equation}
where $L$ is the total number of links in the network, $N$ the total number of nodes, $\sigma_q^2$ the variance of the excess degrees of the nodes, and $A$, $B$, $C$, $D$ are the degrees of the nodes involved in the rewiring step \cite{bertotti2020network}. 
Both $r$ and $K$ are relevant for diffusion processes on the network. Greater assortativity generally leads to an accelerated diffusion in the initial stages \cite{di2023generation}, while a larger $K$ decreases the epidemic threshold (\cite{bertotti2020network} and refs.).
Fig.\ \ref{assort-net} shows an example of a scale-free assortative network obtained with the configuration model followed by rewiring with a target correlation matrix.

The first purpose of this work is to show and test how the assortative scale-free networks generated by our codes can be quickly exported in standard \texttt{graphml} format into the simulation software \texttt{NetLogo}, in order to implement on them the agent-based Bass diffusion model. As recalled in Sect.\ \ref{features}, \texttt{NetLogo} is widely used for agent-based simulations; it has some built-in network structures, but is quite limited under this respect. The Python connector \texttt{PyNetLogo} can be used to control NetLogo and to perform a numerical and statistical analysis of the results of the simulations. In Sect. \ref{ABM-assortative} we report the results of agent-based simulations with assortative networks. Sect.\ \ref{relation-ass-clust} describes a relation occurring, in assortative scale-free networks, between the function $\bar{k}_{nn}(k)$ and the clustering-degree function $C(k)$. A strong correlation was found computing the average of $C(k)$ over an ensemble of networks generated by rewiring. In Sect.\ \ref{review} we review some recent applications of \texttt{NetLogo} to the Bass diffusion model with improved dynamical rules (e.g., including marketing-mix utility functions, analysis of brands, cathegories of adopters, opinion leaders etc.). This review is motivated by the fact that the techniques described in Sect.\ \ref{features} and \ref{ABM-assortative} for assortative networks and for the peak diffusion time could in principle be applied to all these models.

Sect.\ \ref{signed} presents the second main achievement of this work, namely the implementation in \texttt{NetLogo} of the Bass model on arbitrary signed networks, following and extending the work by Mueller and Ramkumar \cite{mueller2023signed}. The extension concerns both the network aspect and the transition rules. Any network which can be generated in \texttt{networkX}, including our assortative scale-free networks and modifications (for example, with cuts in the power law tail), or custom-built small-world networks etc., can be transformed into a signed network with given probability of positive/negative links or also with a continuous random distribution of their weights. Then, any kind of transition rules compatible with \texttt{NetLogo}, like those reviewed in Sect.\ \ref{review}, can be implemented on these networks. Results confirm Mueller's concern that the presence of a threshold for diffusion based on the computation of positive and negative feedback of neighbors hinders the process of spreading by influencing the clustering effect of the network itself. This process is dependent on both the amount and the weight of negative links, provided that there is a random distribution of the seed.

\section{Main features of NetLogo simulations}
\label{features}

\texttt{NetLogo} is a free simulation software for agent-based models, widely used in the social and economic sciences \cite{sklar2007netlogo}. The transition rules which define the behavior of the agents can be set with a high degree of flexibility. We gave in \cite{di2023generation} a first example of agent dynamics corresponding to the classical network Bass model, with a suitable renormalization of the imitation coefficient $q$ in order to allow comparisons among different network topologies.

In the simulation one can distinguish between adoptions due to network-social effects (the Bass $q$-term) and those due to ''broadcast'' effects (the Bass ''advertising'' $p$-term).  The values of the $p$ and $q$ parameters are fixed as global variables, or \texttt{global}. In the setup part of the code the agents are defined, with possible states of adopter or non-adopter. In the network setup part, the procedures \texttt{create-network} and \texttt{make-node} create the new nodes and attach them to the existing nodes according to the prescriptions for Barabasi-Albert networks (only of type-1, i.e., with one link for each new node), Erd\"os-Renyi random networks or simple ring small-world networks. Graphically, a visualization layout is generated. Alternatively, one can import arbitrary networks in \texttt{graphml} format, as explained below. The number of initial adopters (''seed'') can be chosen in the user interface, in addition to the total number of agents, or population, and to the parameters of social influence and broadcast influence.

The \texttt{adopt} procedure for social influence adoptions starts with a count of the number $k$ of nearest neighbors of each node in the network; if the agent on this node has not yet adopted, it adopts with a probability equal to $qk/\langle k \rangle$, where $\langle k \rangle$ is the network average of $k$. The adopters change color in the graphical interface and their total is updated and represented as a function of time (i.e., of the number of steps or ''ticks''), to build a typical S-shaped curve. 

The derivative of the S curve cannot be directly obtained; however, it is possible to manipulate the data produced by the model through the use of extension modules and connectors. Two well known available tools are the \texttt{Mathematica} \cite{bakshy2007netlogo} and \texttt{R} \cite{thiele2012agent} connectors. Here the \texttt{PyNetLogo} connector was used \cite{jaxa2018pynetlogo}, namely a library that allows to control NetLogo by means of the Python programming language. PynetLogo can be called from a main program in order to launch a simulation and to do some straightforward statistical analysis. To this end, one further needs the package \texttt{ctype}, for wrapping the \texttt{dll} libraries and calling functions in shared libraries, and the module \texttt{os}, in order to import the environment variable corresponding to \texttt{JAVA\_HOME}. Also PyNetLogo must be imported and the command \texttt{netlogo = pynetlogo.NetLogoLink(gui=False)} uses the script \texttt{core.py} in PyNetLogo to create a link with NetLogo. Underneath, the NetLogo JVM is accessed through JPype. Once the link with the Java Machine is established, the command \texttt{netlogo.load\_model} allows to open the file with the NetLogo code. The \texttt{netlogo.command} line can be used to start the \texttt{setup} and \texttt{create-network} procedures. One example was shown in \cite{di2023generation} and is available at \texttt{https://github.com/Ladilu/python-bass-accessible}. Then, the \texttt{netlogo.repeat\_report} command is used for returning the desired data (in our case, the number of agents that have adopted) from NetLogo to the Python workspace. This action can be repeated for a certain number of ticks of the simulation. Some basic examples, not related to the current study, are available at \texttt{https://pynetlogo.readthedocs.io/en/latest/}. 

In our case the duration of the simulation has been set to 100 ticks, well beyond the saturation of the diffusion curve. Then the result is transformed into a \texttt{pandas} data-frame and the derivative of the curve is obtained by the \texttt{diff} method. In this way one can identify the adoption peak time $t_{max}$, namely the time at which the number of new adoptions per unit time reaches its maximum. This quantity is useful for characterizing the Bass diffusion process when there are no initial adopters. Such a characterization is not possible in the pure SI model, in which being the $p$-term absent, diffusion does not start without an initial seed. Therefore in the SI model the peak time depends strongly on the initial conditions and cannot be easily put in relation, e.g., to the network topology.

For a random graph the network setup is different. The procedure \texttt{wire4} creates the links with a probability set in the user interface. A window in the user interface displays a list of the nodes' degrees and can be programmed to show the average degree. For importing an external graph, e.g.\ in the case of scale-free assortative networks, an extension of the software is needed called \texttt{nw extension}, with the command \texttt{nw:load-graphml}.
See code examples at \texttt{https://github.com/Ladilu/Mueller\_Bass\_accessible}.

\section{Agent-based simulations with assortative networks}
\label{ABM-assortative}

By performing agent-based simulations of Bass diffusion with the techniques described above, one can obtain information about peak diffusion times, to be compared with the results of models based upon differential equations.

In \cite{di2023generation} we have computed peak diffusion times on scale-free assortative networks generated via configuration model and Newman rewiring. A large system of coupled differential equations was used, namely one equation for each node. This was possible because realizations of the networks were available, and not only their probabilistic degree distribution and correlation functions. It turns out that, unlike in the HMF approximation, the diffusion peak is reached earlier in assortative than in uncorrelated networks. One also observes a strong dependence on the maximum degree effectively present in the network: the larger this degree, the shorter the peak diffusion time.

These results are confirmed by the agent-based simulations: see Tab.\ \ref{tab-1}. It is further observed that when the maximum degree is very large, the peak time is delayed for uncorrelated networks. Note that here the Bass coefficients $p$ and $q$ have been chosen to be smaller than in \cite{di2023generation}, since all agent-based models include a finite time for the spreading process of the technology .

\begin{table}
\begin{tabular}{|F{0.2\textwidth}|T{0.08\textwidth}|T{0.08\textwidth}|T{0.08\textwidth}|T{0.15\textwidth}|T{0.15\textwidth}|T{0.15\textwidth}|}
\hline
\textbf{\hfil Type}&
\textbf{Maximum degree $k_{max}$}&
\textbf{Mean degree $\langle k \rangle$}&
\textbf{Assortativity coeff.\ $r$}&
\textbf{Peak time $t_{max}$ in Python (differential eqs.)}&
\textbf{Maximum adoption rate in Python}&
\textbf{Max.\ rate and time in PyNetLogo (agent-based simuls.)}
\\ 
\hline
\textbf{Assortative} &46&3.8&0.55&6.16&4.86\%&4.63\% at $t=6$ \\
\hline
\textbf{Uncorrelated} &46&3.8&-0.02&7.72&5.4\%&6.14\% at $t=8$ \\  
\hline
\textbf{Assortative}  &92&4.43&0.17&4.84&4.98\% & 6.26\% at $t=6$\\
\hline
\textbf{Uncorrelated}  &92&4.43&-0.09&6.2 &5.59\% & 7.26\% at $t=7$\\
\hline
\textbf{Assortative}  &140&4.32&0.11&4.6&4.99\% & 5.6\% at $t=5$\\
\hline
\textbf{Uncorrelated}  &140&4.32&-0.08&6.32 &5.56\% & 5.8\% at $t=10$\\
\hline
\end{tabular}
\caption{Examples of peak times and maximum values for the derivative of the Bass diffusion curve (''adoption rate''), observed in assortative networks with correlations of the BM1 type \cite{di2023generation} and in uncorrelated networks. All networks have scale-free exponent $\gamma=2.5$, number of nodes $N=1000$, minimum degree $k_{min}=2$ and have been generated via configuration model plus Newman rewiring. The parameters for the Bass diffusion equations are $p=0.025$, $q=0.155$. One observes that the diffusion peak is reached earlier in assortative than in uncorrelated networks.}
\label{tab-1}
\end{table}

\subsection{Relation between assortativity and the clustering-degree function $C(k)$}
\label{relation-ass-clust}

Considerations on the role of clustering in diffusion processes (see also Sect.\ \ref{signed}) suggest to use our algorithm of Newman rewiring for investigating the relation between clustering and assortativity in scale-free networks. This is a further example of the advantage of using full statistical ensembles of networks with pre-defined correlations. Only in such ensembles can certain correlations become evident.

On the relation between clustering and assortativity only a few empirical studies have been published, limited to the total clustering coefficient $C$. For example, in \cite{holme2007exploring} a weak positive correlation between $C$ and $r$ has been reported. Looking at four sets of biological networks, it was found that their representation in an $r$-$C$ plane occupied a bounded region with elongated shape.

\begin{figure}[h]
\centering
\includegraphics[width=.63\textwidth]{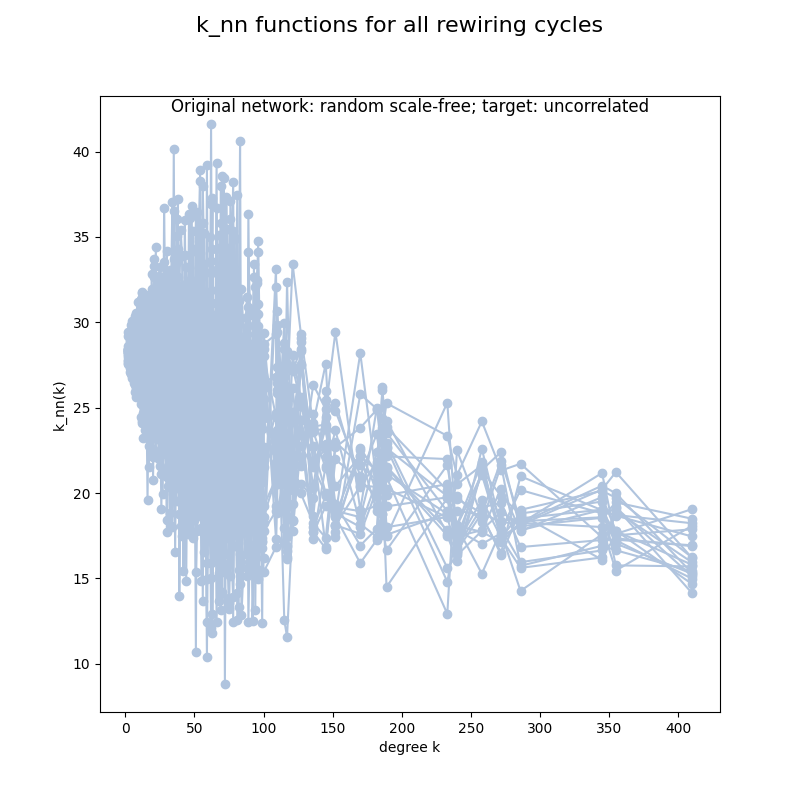}
\includegraphics[width=0.9\textwidth]{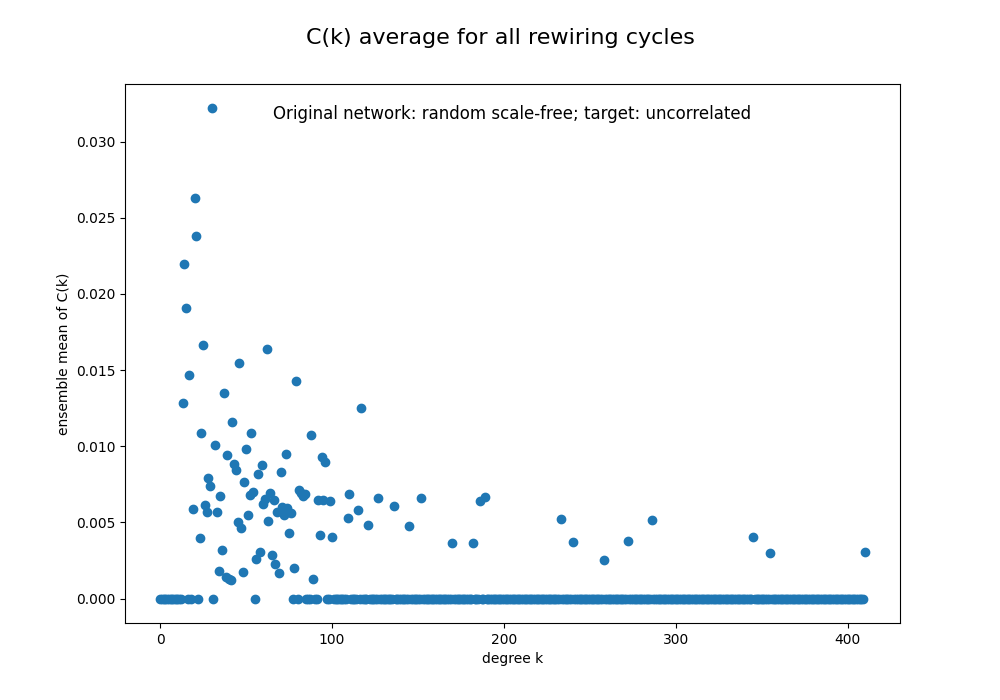}
\caption{Left: cloud of $\bar{k}_{nn}(k)$ functions for a cycle of 20 Newman rewirings with uncorrelated target, starting from a scale-free configuration-model network with $\gamma=2.5$. Right: average of the clustering $C(k)$ as a function of degree over the same ensemble. Networks with 15000 nodes. Approx.\ 130000 rewirings per each cycle. Note the small values of the clustering coefficient at all degrees.
}
\label{knn-C-UNC}
\end{figure}

A qualitative analysis of our assortative networks obtained with Newman rewiring, like that in Fig.\ \ref{assort-net}, indicates that the densely connected core is made of hubs of intermediate size. These nodes have a degree around a value $k_0$ such that the function $\bar{k}_{nn}(k)$ is close to its maximum at $k_0$; they are thus able to find in the network approximately $k_0$ neighbors of similar degree. On the contrary, the hubs with $k \gg k_0$ cannot find enough assortative partners with similar degree, and therefore lie on the decreasing part of the $\bar{k}_{nn}(k)$ plot. It can be expected that the assortative core has a high clustering level, and this can be confirmed by computing the average over several rewiring cycles of the function $C(k)$ which gives the mean clustering coefficient of the nodes of the newtwork having degree $k$.

\begin{figure}[h]
\centering
\includegraphics[width=.63\textwidth]{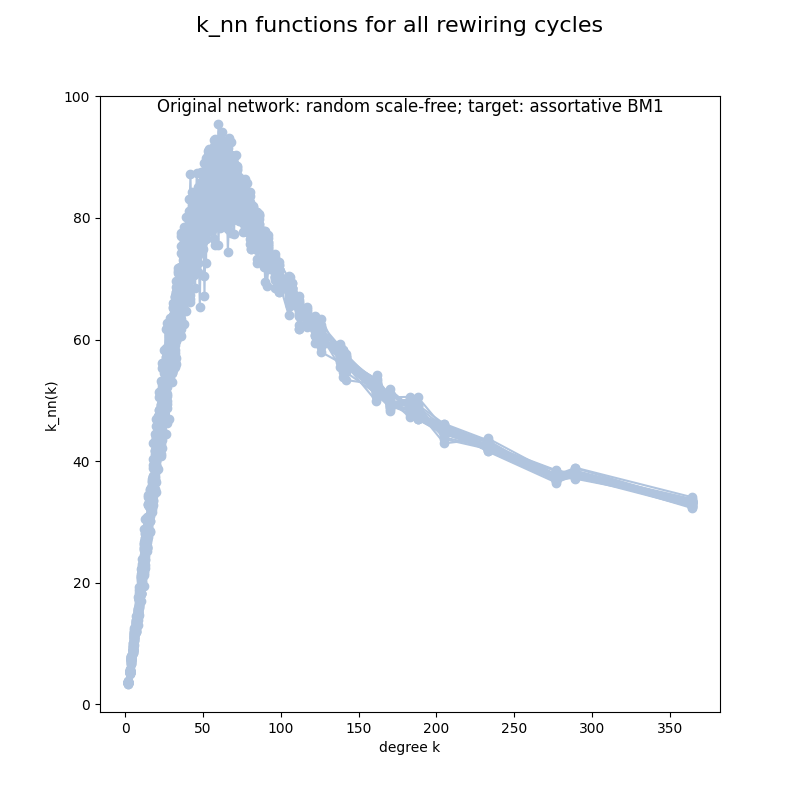}
\includegraphics[width=0.9\textwidth]{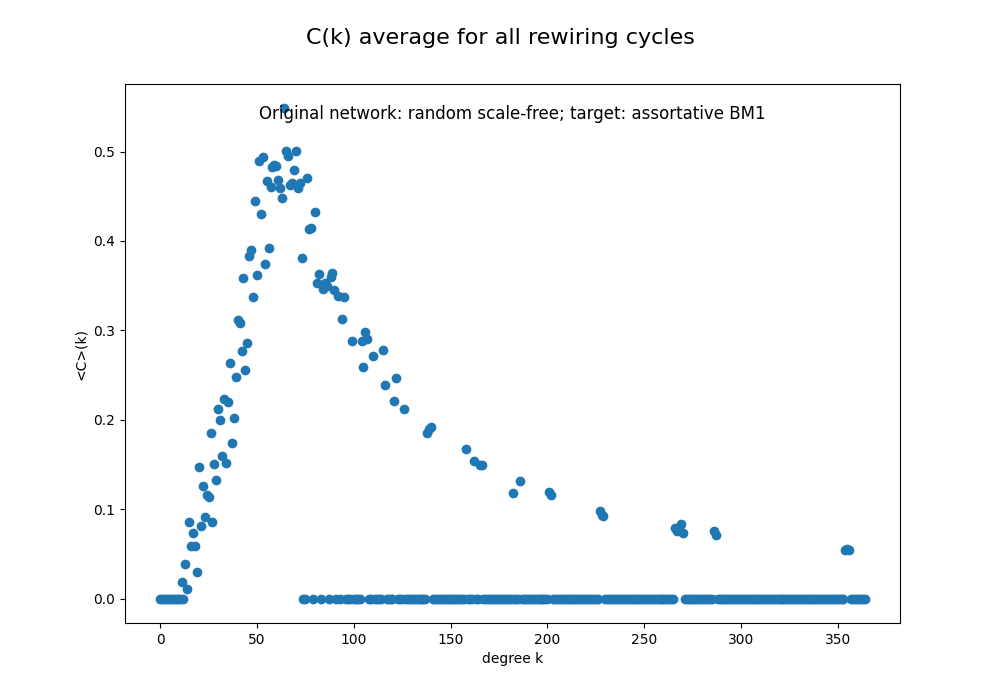}
\caption{Left: cloud of $\bar{k}_{nn}(k)$ functions for a cycle of 20 Newman rewirings with assortative target, starting from a scale-free configuration-model network with $\gamma=2.5$. Right: average of the clustering $C(k)$ as a function of degree over the same ensemble. Note the high level of maximum clustering in the assortative core of the network and the strong correlation between the curves $\bar{k}_{nn}(k)$ and $C(k)$. Networks with 15000 nodes. Approx.\ 130000 rewirings per each cycle.
}
\label{knn-C-ASS}
\end{figure}

The function $C(k)$ has been previously evaluated for some large real social networks \cite{hardiman2009calculating} and for a class of rewired BA networks \cite{tam2008construction}, but is otherwise little known. We find that in a statistical ensemble of uncorrelated scale-free networks built via configuration model, $C(k)$ is very small and exhibits a random behavior (see Fig.\ \ref{knn-C-UNC}, obtained through Newman rewiring with uncorrelated target). Using a correlated target, the function changes completely and clearly shows a high level of clustering in the assortative core, see Fig.\ \ref{knn-C-ASS}. Codes are available at \texttt{https://github.com/Ladilu/Mueller\_Bass\_accessible}.

\subsection{Other agent-based models of Bass diffusion}
\label{review}

This Subsection contains a brief review of recent results on agent-based simulations of the Bass diffusion model. The techniques described above for assortative networks and for the peak diffusion time could be applied to these models. In Sect.\ \ref{signed}, however, we will focus on the version proposed by Mueller and Ramkumar \cite{mueller2023signed} on signed networks, which requires some specific steps for assigning the weights of the graph edges. 

Kiesling et al.\ \cite{kiesling2012agent} describe several applications of agent-based Bass diffusion models. They can serve either as intuition aids for the purpose of theory building or as tools to analyze real-world scenarios and obtain policy recommendations, also in sight of future management decisions which need some form of support. The latter stream of research integrates the Bass model with further elements which were not introduced at first in \cite{bass1969new}, namely the marketing mix factors: product, price, promotion and distribution. 

An analysis of brands was implemented in \cite{schramm2010agent}. The adopters are divided in three categories: pioneers, early adopters and late adopters. The model combines brand-level diffusion curves in order to form a product-type diffusion curve. In order to know the marketing mix parameters for the brands it is enough to have an initial evaluation of the brand agents' characteristics and of the proportion of the adopters. Then, an individual threshold can be built at the microlevel for simulating with NetLogo the diffusion at the macro level of a given product category. In this real-world scenario the network is fixed but in principle an extension to heterogenous networks is allowed and encouraged.

Another simulation dynamics is based on the concept of critical mass for diffusion, introduced systematically by Watts and Dodds \cite{watts2007influentials} as constituted by individuals who are easily influenced by ''influentials'', namely the ones who contribute to create the determinant opinion about a single product. The influentials came to occupy a central place in the literature dealing with diffusion of innovations, of which one main example is the previous 1995 edition of \cite{rogers2010diffusion}. They are also called opinion leaders and they are especially relevant to the diffusion of certain technologies such as mobile phones and computers, which change the social behaviour of people. 

One study which examines the opinion leaders' role by means of agent-based simulations is \cite{van2011opinion}. It is found that innovative behaviour of opinion leaders yields a higher adoption percentage; in addition to that, another advantage of the presence of opinion leaders is a faster diffusion process, as it was measured from simulations. The threshold for adoption is calculated using a simplified version of the model by \cite{delre2007diffusion}. In the latter study, the probability of adoption for an agent $i$ is given by
\begin{equation}
a_{i,j}=P(U_{i,j} \geq U_{i,MIN})
\end{equation}
where
\begin{equation}
U_{i,j}= \beta_{j} x_{i} + (1-\beta_{j}) y_{i}
\end{equation}
\begin{equation}
\centering
y_{i} =\left\{\begin{split}
q_{j}\geq p_{i} \rightarrow{} 1\\\
\textit{otherwise} \rightarrow 0\\ 
\end{split}\right.
 \end{equation}
\begin{equation}
\centering
x_{i} =\left\{\begin{split}
A_{i}\geq h_{i} \rightarrow{} 1\\\
\textit{otherwise} \rightarrow 0\\ 
\end{split}\right.
 \end{equation}
Here, $U_{i,j}$ is the utility agent $i$ has if it adopts innovation $j$ and $U_{i,MIN}$ specifies its minimum utility requirement. The individual preference $y_{i}$ and the local social influence $x_{i}$ on the agent $i$ are both threshold functions, whereas $\beta_{j}$ represents how strong the social influence effect is in the market of the product $j$.   In the definition of the social influence part, $p_{i}$ is the individual preference of agent $i$ and $q_{j}$ is the quality of innovation $j$. In the definition of the individual part, $h_{i}$ is the threshold value for the fraction of adopters in $i$'s personal network, namely the value that must be overcome in order to activate adoption for agent $i$. $A_{i}$ is the fraction of adopters in the $L$-th order set of alters of agent $i$ -- the alters are the agents included in $i$'s personal network. Both $L=1$ and $L=2$ were examined in \cite{delre2007diffusion}. In a later work, \cite{delre2007targeting}, the local social influence was re-defined as
\begin{equation}
\centering
x_{i} =\left\{\begin{split}
a_{i}\geq h_{i} \rightarrow{} 1\\\
\textit{otherwise} \rightarrow 0\\ 
\end{split}\right.
 \end{equation}
 \begin{equation}
 a_{i} = \frac{\textit{adopters}_{i}}{\textit{adopters}_{i} + \textit{non-adopters}_{i}}
 \end{equation}
 where $a_{i}$ is the percentage of adopters in $i$'s personal network.

In the model by Bohlmann et al.\ \cite{bohlmann2010effects}, a threshold defines the minimum proportion of adopters among all those connected to individuals before they can adopt. The simulation is carried out with 1600 agents, of which 50 are selected randomly as innovators who have adopted at the earliest stage of the innovation. This external influence is fixed consistently among all topologies of the network. The agent examines the status of the immediate neighborhood; if the percentage of the adopters exceeds the threshold, the agent adopts with probability 0.5. Four topologies are selected: cellular automata, random network, small world network (random and small world use Poisson distribution, their average number of links per vertex are the same), power-law network (BA network). Five levels of the adoption threshold are allowed: 0,01, 0.05, 0.10, 0.15, 0.20. The peak time is compared for the different topologies and a threshold of 0.05. All simulations show a diffusion cascade for the lower values of the threshold, the percentage diminishes from 100\% to less than 100\% when the threshold reaches the values of 0.15 and 0.20. The adoption threshold negatively affects the number of new adopters at the point of peak adoption. The effects are less pronounced for the random network under diffusion cascade. The network topology has a more pronounced effect on diffusion processes for higher adoption thresholds. At a high threshold level of 0.20, the random network fails to diffuse at all, whereas the other networks have pronounced differences in their peak adoption times. A model in which the agents belong to one of two market segments is proposed for a first analysis of the peak of the adoption and is foreseen as a future application for agent-based simulations of an heterogeneous market configuration.

In the work of Delre et al.\ \cite{delre2010will}, the graph is fully connected, the percentage of initial adopters is fixed and they are randomly distributed. In contrast to percolation models without social influence, in this model it is possible that an agent first does not adopt when being informed about the product, but later, when several neighbors have adopted, he or she may decide to adopt as well because of the increased social utility of the product. If agent $i$ is informed about product $g$ and he or she decides to adopt it, he or she is considered an adopter until the end of the simulation. For the purpose of choosing a suitable network configuration a more realistic version of the scale-free network (Amaral et al., \cite{amaral2000classes}) is adopted. Here, when a new node is attached to the network, the probability of all the other nodes of being selected for the attachment is still proportional to the number of nodes they already have, but it decays exponentially due to a fixed probability $h$ to become inactive at any moment of the process. The scale-free network of Amaral et al.\ also yields a power law distribution of links for low connected links, but the number of links decays faster when the probability $h$ increases. In other words, the scale-free network model is generalized as follows: Vertices are classified into one of two groups, the active and the inactive. All vertices are created active but in time they can become inactive, namely they cannot receive any more links. The constraints that are responsible for the transition of the vertices from active to inactive lead to cutoffs on the power law decay of the tail of connectivity distribution and when they are strong enough no power law region is visible.

It is important to mention that transitions between nodes' different states have been modeled in SIR epidemic diffusion studies by means of Markov processes, which actually allow to go beyond the mean field model -- see for example the GEMF model in \cite{sahneh2013generalized} -- with a time-based reconstruction of the transient region and the introduction of the study of metastable states. It was correctly stated in \cite{banisch2012agent} that this approach is based on time correlations and differs from the one of agent-based models, which actually work on cross-correlations and therefore generates the final state from the transient by means of interaction. The differential equations that must be solved in the case of a Markov process are generally yielding a space state far too big both for analytical and numerical solutions; on the other side, Markov chains analysis and ABMs techniques do not overlap in a simplified way. In other words, it is not possible to directly reconstruct the state of the agents form their memory data but statistical analysis has to be performed. We believe that usage of ABMs of previously constructed large networks constitutes a valid alternative to real datasets analysed by means of Markov Chains, within the limits offered by the computational constraints. The usage of the Python language allows to re-create large graphs that would take an excessive amount of time in NetLogo and therefore allows for a better numerical treatment of the problem.

\section{The Bass model on signed networks}
\label{signed}

In their work of 2023 \cite{mueller2023signed}, Mueller and Ramkumar note that most models of innovation diffusion assume that relationships in a network are of a positive nature, in the sense that adoption by an agent generates a probability of adoption by the agents connected to it. Some authors, however, have also considered inverted influence effects in the form of negative word of mouth or foe relationships. In terms of the network, these approaches involve so-called signed graphs or nets, friendship-foe relationship networks or networks with negative weights. Earlier works involving signed networks are cited in \cite{mueller2023signed} and mainly concern Heider's model of balanced sentiments in social networks and opinion dynamics models. 

Barbuto et al.\ \cite{barbuto2019improving} use a practical definition of diffusion of innovation based on \cite{rogers2010diffusion} and social network analysis. Taking into account the existing literature \cite{banerjee2013diffusion} based on centrality statistics and a regression model, they construct an adapted linear regression analysis which requires a choice of coefficients and the consideration of some parameters like degree centrality, betweenness, clustering coefficient. Barbuto et al.\  consider how the literature shows that one can obtain different diffusion rates by changing the so-called injection points (IP), or early adopters distribution. They perform agent-based model simulations with NetLogo in which the agents who have adopted reconsider their preference at every iteration step, namely they receive information from neighbors, pass information to neighbors and finally decide to adopt if the preference for the new technology is greater than a fixed adoption threshold. In order to augment the model reality and in sight of practical applications with empirical data, they introduce the disappointment extension, which includes the possibility that the innovation disappoints the adopter. When
this case occurs, the model provides a possible third action for the ordinary agent, to reduce its preference towards the innovation by a variable amount (between 25 and 75\%). In this version of the model, the disappointment is modeled as a random event which affects a variable fraction of the population (between 0 and 25$\%$). The results of the ABM simulations indicate that the individual properties of the IPs are not primarily responsible for the diffusion process. Instead, the density and clusterization of the network play a significant role.

The model proposed in \cite{mueller2023signed} for agent-based simulation employs a network (first random, then small-world or BA) with $N$ agents and $L$ links where the sign of existing links are randomly switched from positive to negative, with a probability $p_n$. This random distribution of negative links in the network is in contrast with balance theory, which assumes that some triangles in signed networks are likely to be found. 

It is then assumed that an agent $i$ adopts the innovation at each timestep according to a threshold condition, namely if the sum $(\theta_- \cdot f_{FOE} + \theta_+)$ is strictly positive. Here $\theta_-$ represents the number of first neighbors who have already adopted and are connected to the agent through a negative link, and similarly for $\theta_+$. The coefficient $f_{FOE}$ describes the strength of the negative influence and typically varies between -3 and 0; the value -1 represents the case in which a negative link has the same strength as a positive one. When $f_{FOE}$ is a simple fraction, an interesting step-wise foe effect is observed in the final adoption rate; for example, if $f_{FOE}=-0.5$, then it takes exactly 2 negative links to counter the effect of a positive link, ans therefore for $f_{FOE}$ slightly larger than that value the adoption rate has a positive jump.

Simulations are performed with a custom-made code (not available), varying mainly the parameters $p_n$, $f_{FOE}$ and the kind of network. The starting scenario is a random network with 300 nodes, 900 links and a population of 3\% initial adopters randomly chosen. The principal output of the simulations is the final percentage of adopters. While in the traditional Bass model the entire population eventually adopts the innovation, here the presence of a threshold has the consequence that a certain number of individuals never adopt, especially if their neighbors are connected with negative links. These ''resistant'' nodes also become an obstacle for the adoption of their contacts with positive links, because they make it harder to reach the threshold. The whole process depends strongly on the parameters $p_n$ and $f_{FOE}$. In fact, the curves displaying the final adoption rate as a function of $p_n$ (''share negative ties'') always have a decreasing behavior, like an inverted S-curve passing from 100\% final adoptions when $p_n=0$ to approx.\ 0\% when $p_n=1$.

The reference curve with $f_{FOE}=0$ is markedly higher than the curve with, e.g., $f_{FOE}=-1$: in the middle of the graph, with $p_n=0.5$, the observed difference is about 40\% in the final adoption rate. For comparison, the differences in the adoption rate due to the effect of the network for equal $p_n$ and $f_{FOE}$ are at most a few percent. They are more significant at large $p_n$, as relative value, when the final adoption rate is small. In addition to random networks, BA and small-world networks are employed. Their exact features are not specified in the paper. It is recalled that, as general criteria, a high clustering level of the network facilitates diffusion in the case of complex contagion, while a small average path length and a fat tail in the degree distribution are beneficial for simple contagion processes \cite{centola2007cascade}, \cite{kiesling2012agent}, \cite{delre2007diffusion},\cite{ramkumar2022diffusion},\cite{tur2018diffusion}[...]. In the presence of many negative links one expects for BA networks several adoption blockades in the network periphery, where a single resistant node can block many others, due to the low clustering. On the other hand, in small-world networks the regions with high clustering are more frequently blocked. In order to confirm these hypotheses, Mueller and Ramkumar produce graphs which relate the nodes' degrees to their probability of being a final non-adopter.

As anticipated in the Introduction, in this work we reproduce and extend the model of \cite{mueller2023signed} by applying it to a wide set of networks generated with networkX using the configuration model and Newman rewiring. These networks are imported into NetLogo, and then NetLogo performs the agent-based simulations using the threshold transition rule of \cite{mueller2023signed}. In order to assign negative signs to some links with a pre-defined probability $p_n$, we use the following trick. NetLogo allows to assign to each link a random weight with uniform distribution in a user-defined interval. It is easy to check that if the interval is equal to $(-\alpha,1)$, where $\alpha=p_n/(1-p_n)$, then the total probability to obtain a negative weight is actually $p_n$. Since in the transition rule only the sign of the link matters, and not its weight, this implementation is equivalent to that of \cite{mueller2023signed}, further allowing to simulate, if needed, a more general situation in which the weights of the links play a role. In other words, one could simulate a model in which links can influence diffusion positively or negatively with a continuous distribution of strengths.

\subsection{Parallel runs of NetLogo simulations}

\begin{figure}[ht] 
\centering \includegraphics[width=0.6\columnwidth]{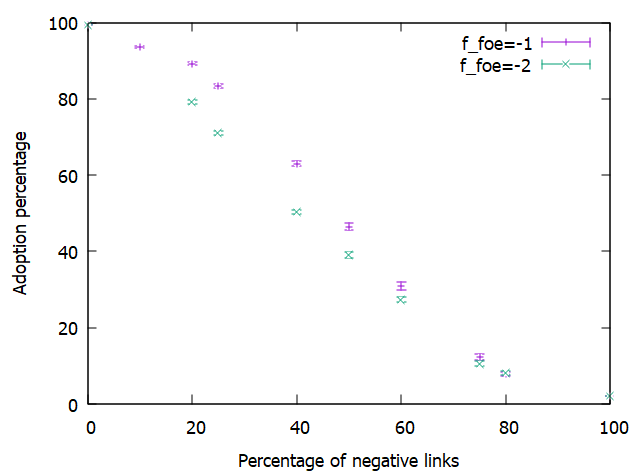}
\caption{
Plot of the two curves that can be obtained for the values $f_{FOE}= -1$, $f_{FOE}= -2$  in $(\theta_- \cdot f_{FOE} + \theta_+)$ according to the Mueller-Ramkumar scheme for the threshold needed for adoption. On the $x$-axis the percentage of negative links in the network is represented, on the $y$-axis the percentage of the number of adoptions can be found.
}
\label{plot_two_values_ffoe}
\end{figure}

The already mentioned tool named PyNetLogo for connecting NetLogo with a Python environment can also be used in order to run a set of simulations and collect the desired statistics. This feature is generally described in \texttt{https://pynetlogo.readthedocs.io/en/latest/}; in that context the \texttt{ipyparallel} Python package is proposed, for the purpose of running on 4 clusters the subsequent simulations. The document also describes the use of \texttt{pandas} data-frames for further analysis. In the current problem it was found that actually the package \texttt{concurrent\_futures} better satisfies our needs of fast and secure parallelization of the tasks, with the number of \texttt{max\_workers} of \texttt{PoolExecutor} set to 4. This allows to exploit the 4 processors that we have currently available. It takes a few tens of seconds to run 100 simulations, then the analysis must be performed separately by means of data-frame statistics and writing to files for plotting with Gnuplot. For our purposes, extracting values for the maximum adoption rate and calculating the mean and standard deviation of each  data-frame maximum are the desired statistical output. In this way two curves have been reconstructed for the study of the Mueller-Ramkumar model, using a random network with 3$\%$ initial seed for diffusion and setting two values of the parameter $f_{FOE}$, namely $-1$ and $-2$. The two datasets show the behaviour of the maximum adoption rate as a function of the percentage of negative links -- see Fig.\ \ref{plot_two_values_ffoe}.

\subsection{Results with Mueller-Ramkumar threshold model on assortative and uncorrelated networks}

The set of 100 runs of Netlogo achieved by means of parallel computing with PyNetLogo and the \texttt{concurrent\_future} Python package was repeated in order to obtain more statistics about the effect of Mueller threshold on the diffusion on an assortative network generated by means of the \texttt{configuration model} function and Newman rewiring with BM1. The random negative links are added with $50\%$ probability in NetLogo after the \texttt{graphml} format of the network is loaded and the results are compared with the ones obtained by means of an analogous procedure but starting from an uncorrelated network. The output for different maximum degrees from a start configuration of 500 nodes and a scale law with $\gamma=2.5$ are presented in Tab.\ \ref{tab2}. It is visible how the diffusion is higher in the uncorrelated case; the values are demonstrated to be stable by the size of the standard deviation, which is compatible with a detailed analysis.
\begin{table}[htp]
\begin{tabular}{|F{0.2\textwidth}|T{0.1\textwidth}|T{0.1\textwidth}|T{0.1\textwidth}|T{0.1\textwidth}|T{0.1\textwidth}|T{0.3\textwidth}|}
\hline
\textbf{\hfil Type}&
\textbf{Maximum degree $k_{max}$}&
\textbf{Mean degree $\langle k \rangle$}&
\textbf{Assortativity coefficient $r$}&
\textbf{Mean\, of\, diffusion\, in\, PyNetLogo}&
\textbf{Stand.\, deviation of the mean}
\\ 
\hline
\textbf{Assortative} &31&3.66&0.51&29.53\%&0.31 \\
\hline
\textbf{Uncorrelated} &31&3,66&-0.01&39.06\%&0.37 \\  
\hline
\textbf{Assortative}  &56&4.25&0.28&32.83\% &0.35 \\
\hline
\textbf{Uncorrelated}  &56&4.25&-0.04&41.87\% &0.39 \\
\hline
\textbf{Assortative}  &79&4.3&0.13&33.15\% &0.33 \\
\hline
\textbf{Uncorrelated}  &79&4.3&-0.08&41.27\% &0.38 \\
\hline
\end{tabular}
\caption{Examples of diffusion values with the Mueller threshold as it is observed in assortative networks with correlations of the BM1 type, $\gamma=2.5$, $N=500$, $k_{min}=2$ and in uncorrelated networks. The assortative networks have been generated via configuration model plus Newman rewiring.}
\label{tab2}
\end{table}

\begin{figure}[p]
\begin{subfigure}[H]{\linewidth}
        \includegraphics[width=125mm]{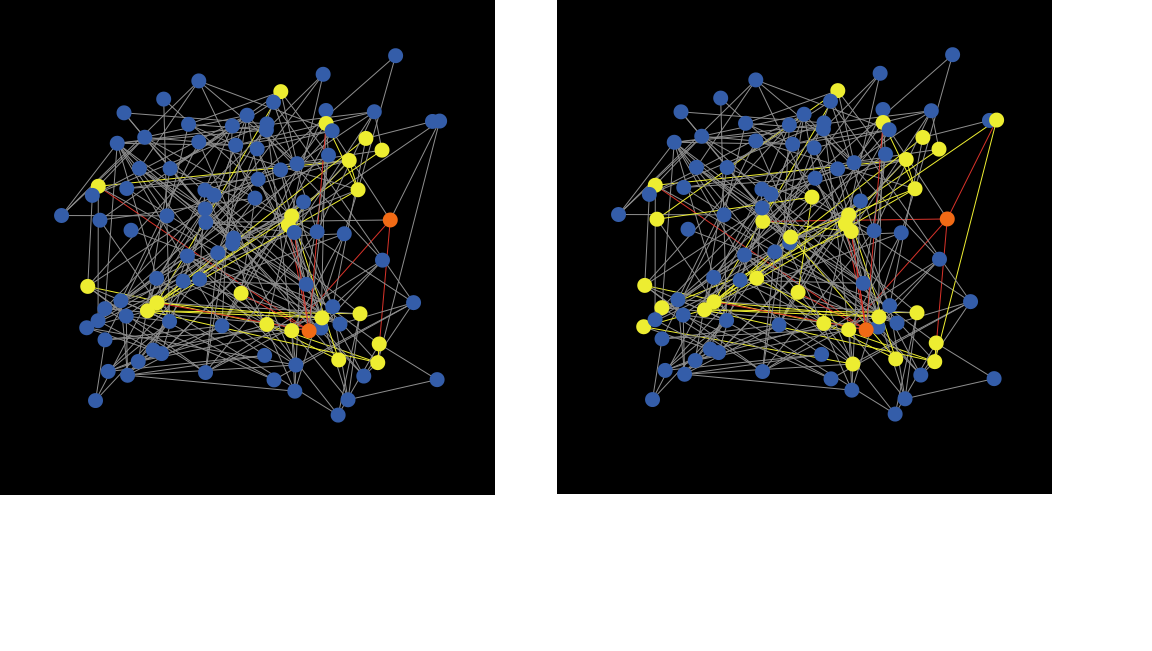}
        \label{fig:a}
\end{subfigure}
\begin{subfigure}[H]{\linewidth}
        \includegraphics[width=125mm]{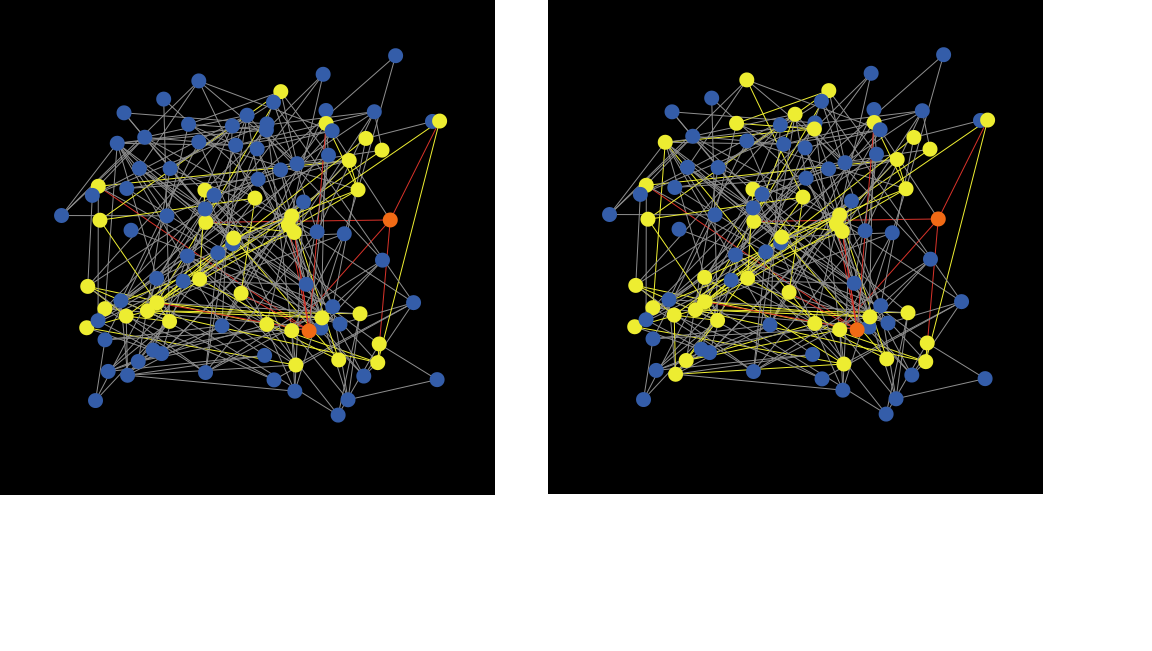}
        \label{fig:b}
\end{subfigure}
\caption{Some \texttt{NetLogo} views of the agent dynamics for adoption with the Mueller  threshold and $f_{foe} =-1 $ in a simplified random network. The agents that have not adopted are initially colored in blue, the seed agents are highlighted in orange, adoptions are represented by  yellow color. The network has 100 nodes and wiring probability 0.05. }
\label{fig:NetLogo}
\end{figure}

\section{Conclusions}
\label{concl}

In this article a new framework for diffusion studies on networks was proposed, based on numerical codes written in \texttt{Python-networkX} and \texttt{NetLogo}. Although these free software packages are well-known and widely employed, their combined use is not simple (\texttt{networkX} is for network generation and analysis, \texttt{NetLogo} for agent-based simulations). A connector tool called \texttt{PyNetLogo} is available and partially documented; it also allows to run parallel simulations via the \texttt{ipyparallel} Python package. There are no previous examples in the literature, however, of the application of these methods to established diffusion models which have been defined and analyzed in the past with systems of differential equations. Besides the specific models considered in this work (Bass model and Mueller-Ramkumar threshold model), several other models can potentially benefit from agent-based simulations on networks. Some of them have been briefly reviewed in Sect.\ \ref{review}.

In our work we have performed an analysis of the emerging features of dynamics of preference diffusion about technology, based on ABM simulations. We recall that both Markov Chain Approaches (MCA) and Agent Based Models (ABM) have an empirical usage and the most general subject where both lines ground their common root is structure (or pattern) generation. While in the context of MCA, patterns emerge from time correlations or memory effects in the lifetime of a system, in the context of ABM the emerging structures are mostly related to cross-correlations between the agents. We have focused on the latest approach, providing a first statistical analysis of the main emerging quantitative characteristic features, namely the diffusion maximum and the peak time for diffusion. We believe that also opinion models with other interaction mechanisms could benefit from a descriptive representation given through our scheme, constituted by an input network provided by Python rewiring and NetLogo simulations for the diffusion on these imported graphs. 

The main practical results are summarized in Tab.\ \ref{tab-1}, where diffusion times are compared for assortative and uncorrelated networks and differential equation results are put together with the agent-based approach outcome. The findings for Mueller's model are presented in Tab.\ \ref{tab2}, for both assortative and uncorrelated networks. This is useful for social network analysis, because social networks are typically assortative. The uncorrelated networks represents the most simplified model for diffusion and therefore constitute a baseline case for a relevant comparative approach.

\bibliographystyle{ieeetr}
\bibliography{nets}

\end{document}